\documentstyle[amssymb,epsf,cite,11pt]{article}

\newcommand{\be}{\begin{equation}}
\newcommand{\ee}{\end{equation}}

\begin{document}

\title{\bf Fractal and multifractal properties of a family of fractal networks }

\author{ Bao-Gen Li$^{1}$, Zu-Guo Yu$^{1,2}$\thanks{
  Corresponding author, email: yuzg1970@yahoo.com}  ~and Yu Zhou$^{1}$\\
{\small$^1$ Hunan Key Laboratory for Computation and Simulation in
Science and Engineering and }\\
{\small Key Laboratory of Intelligent Computing and Information
Processing of Ministry of Education,}\\
{\small Xiangtan University, Xiangtan,  Hunan 411105, China.}\\
{\small $^{2}$School of Mathematical Sciences, Queensland University of Technology,}\\
{\small GPO Box 2434, Brisbane, Q4001, Australia.}
 }
\date{}
\maketitle

\begin{abstract}
In this work, we study the fractal and multifractal properties of
a family of fractal networks introduced by Gallos {\it et al.} (
{\it Proc. Natl. Acad. Sci. U.S.A.}, 2007, {\bf 104}: 7746). In
this fractal network model, there is a parameter $e$ which is
between $0$ and $1$, and allows for tuning the level of fractality
in the network. Here we examine the multifractal behavior of these
networks, dependence relationship of fractal dimension and the
multifractal parameters on the parameter $e$. First, we find that
the empirical fractal dimensions of these networks obtained by our
program coincide with the theoretical formula given by Song {\it
et al.} ( {\it Nat. Phys}, 2006, {\bf 2}: 275). Then from the
shape of the $\tau(q)$ and $D(q)$ curves, we find the existence of
multifractality in these networks. Last, we find that there exists
a linear relationship between the average information dimension
$<D(1)>$ and the parameter $e$.
\end{abstract}



\section{Introduction}
\ \ \ \ Complex networks have caused extensive attention due to
their close connection with so many real-world systems, such as
the world-wide web, the internet, energy landscapes, and
biological and social systems~\cite{Song2005}. The fractality and
percolation transition~\cite{Rozenfeld2009}, fractal
transition~\cite{Rozenfeld2010} in complex networks, and
properties of a scale-free Koch networks~\cite{Liu2010, Zhang2009,
Zhang2008} have turned to be hot topics in recent years.

Fractal analysis (using the fractal dimension) is a useful method
 to describe global properties of complex fractal
sets~\cite{Mandelbrot1983, Feder1988, Falconer1997}. Song {\it et
al.}~\cite{Song2005, Song2007} proposed an algorithm to calculate
the fractal dimension of complex networks which can unfold their
self-similar property. They mentioned that the box counting
fractal analysis is an effective tool for the further study of
complex networks. But the fractal analysis is not enough when the
object studied can not be described by a single fractal dimension.
It has been found that the multifractal analysis (MFA) is a
powerful tool in both the theory and practice to describe the
spatial heterogeneity of fractal object
systematically~\cite{Grassberger1983, Halsey1986}. The MFA was
originally raised to handle turbulence data, and now it has been
successfully applied in many fields, such as financial
modelling~\cite{Canessa2000, Anh2000}, biological
systems~\cite{Yu2001a, Yu2001b, Yu2003, Yu2004, Yu2006, Yu2010b,
Anh2002, Zhou2005, Han2010, Zhu2011} and geophysical
systems~\cite{Kantelhardt2006, Veneziano2006, Venugopal2006,
Yu2007, Yu2009, Yu2010a, Zang2007}. Lee and Jung~\cite{Lee2006}
found that MFA is the best tool to describe the probability
distribution of the clustering coefficient of a complex network.
Furuya and Yakubo \cite{Furuya2011} analytically and numerically
demonstrated the possibility that the fractal property of a
scale-free network cannot be characterized by a single fractal
dimension when the network takes a multifractal structure. Almost
at the same time, Wang {\it et al.} ~\cite{Wang2012} proposed a
modified fixed-size box-counting algorithm to study the
multifractal property of complex networks.

In this paper, we study the fractal and multifractal properties of
a family of complex networks introduced by Gallos {\it et
al.}~\cite{Gallos2007}. In order to imitate the fractal property
of many scale-free networks found in nature, Song {\it et al.}
\cite{Song2006} developed a network model to describe the
fractality of networks. The main characteristic of this model is
the introduction of a parameter $e$ which could be used to control
the original hubs whether continue to form connections between the
nodes in the process of the growth of complex networks. The
authors \cite{Song2006} pointed out that the parameter $e$ can be
regarded as a level of fractality of the network. The network
corresponds to a pure fractal network which is a pure fractal set
(defined by Mandelbrot \cite{Mandelbrot1983}) when $e = 0$, and a
pure small world network when $e = 1$~\cite{Rozenfeld2009}. Later
on, Gallos {\it et al.}~\cite{Gallos2007} proposed a generalized
version of this network model.

In Section 2, we introduce the generalized version of the network
model in Ref. \cite{Gallos2007} and some of its topological
properties. In Section 3, we examine the fractal dimension of
these networks . A new fixed-size box-counting algorithm for MFA
of networks modified from the one proposed in Ref. \cite{Wang2012}
is given in Section 4. The multifractal properties of the model
networks and their results are also given in this Section. Some
conclusions are presented in Section 5.

\section{Network model}
\ \ \ \ A graph (or network) is a collection of nodes which denote
the elements of a system, and links or edges which identify the
relations or interactions among these elements. In this section,
the algorithm of the generalized version of the network model in
Ref. \cite{Gallos2007} is presented. The network could be obtained
by a method described as follows.

First we give a real number $0\le e \le 1$, and two positive
integers $m$ and $x$ ($x\le m$). In the generation $n=0$, we start
with only two nodes and one edge between them. In order to get the
network of the generation $n+1$, every endpoint of each edge $L$
in the network of the generation $n$ is attached to $m$ new nodes.
Then we generate a random number $p$ which obeys the uniform
distribution between 0 and 1. If $0\leq p<e$, each edge $L$ of the
generation $n$ is kept and $x-1$ new edges are appended to connect
pairs of the new nodes attached to the endpoints of $L$;
otherwise, for each edge $L$ of the generation $n$, we add $x$ new
edges matching new nodes at the ends of $L$ and remove $L$ (see
Fig. 1). As shown in Fig. 2, if we take $e=0,m=2,x=2$,  we add
$m=2$ new nodes to the two endpoints of the sole edge in the
generation $n=0$ to get the network of the generation $n=1$. Due
to $e=0$, we take away the edge of $n=0$ and add $x=2$ edges
between the new nodes. Notice that when $x=1$, we get tree
structure without loop for any value of $e$.

\begin{figure}[!tbp]
\centering \epsfxsize=11cm \epsfbox{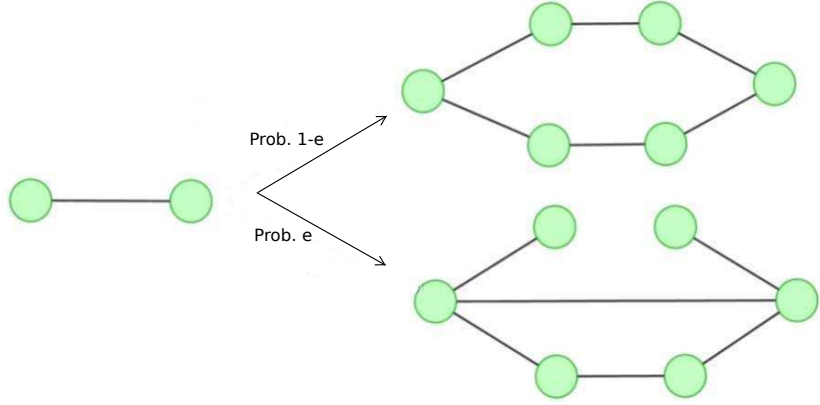}
\caption{Construction of network. The link between hub remains
with probability $e$, otherwise, it is replaced by another link
between new nodes with probability $1-e$.}\label{1}
\end{figure}

\begin{figure}[!htp]
\centerline{\epsfxsize=5cm \epsfbox{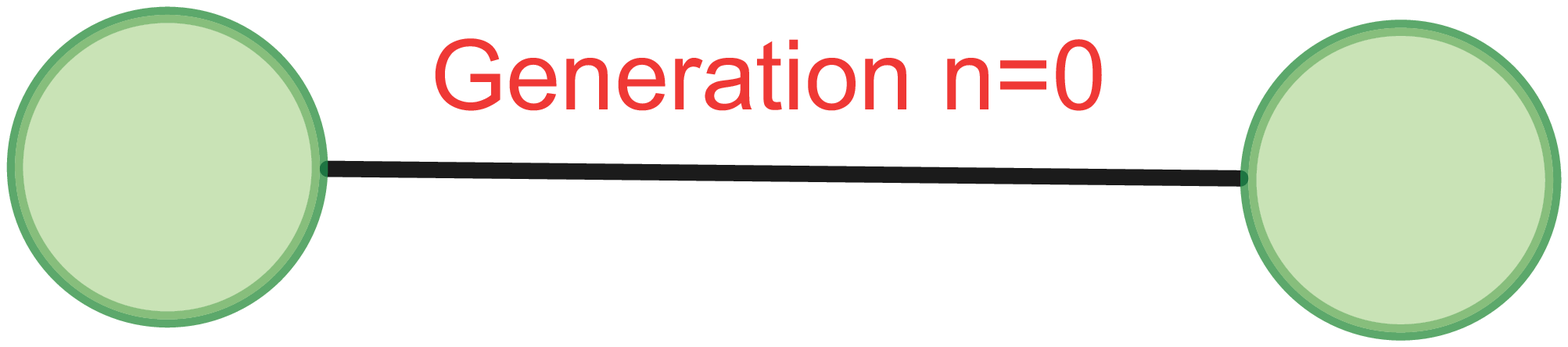} \epsfxsize=5cm
\epsfbox{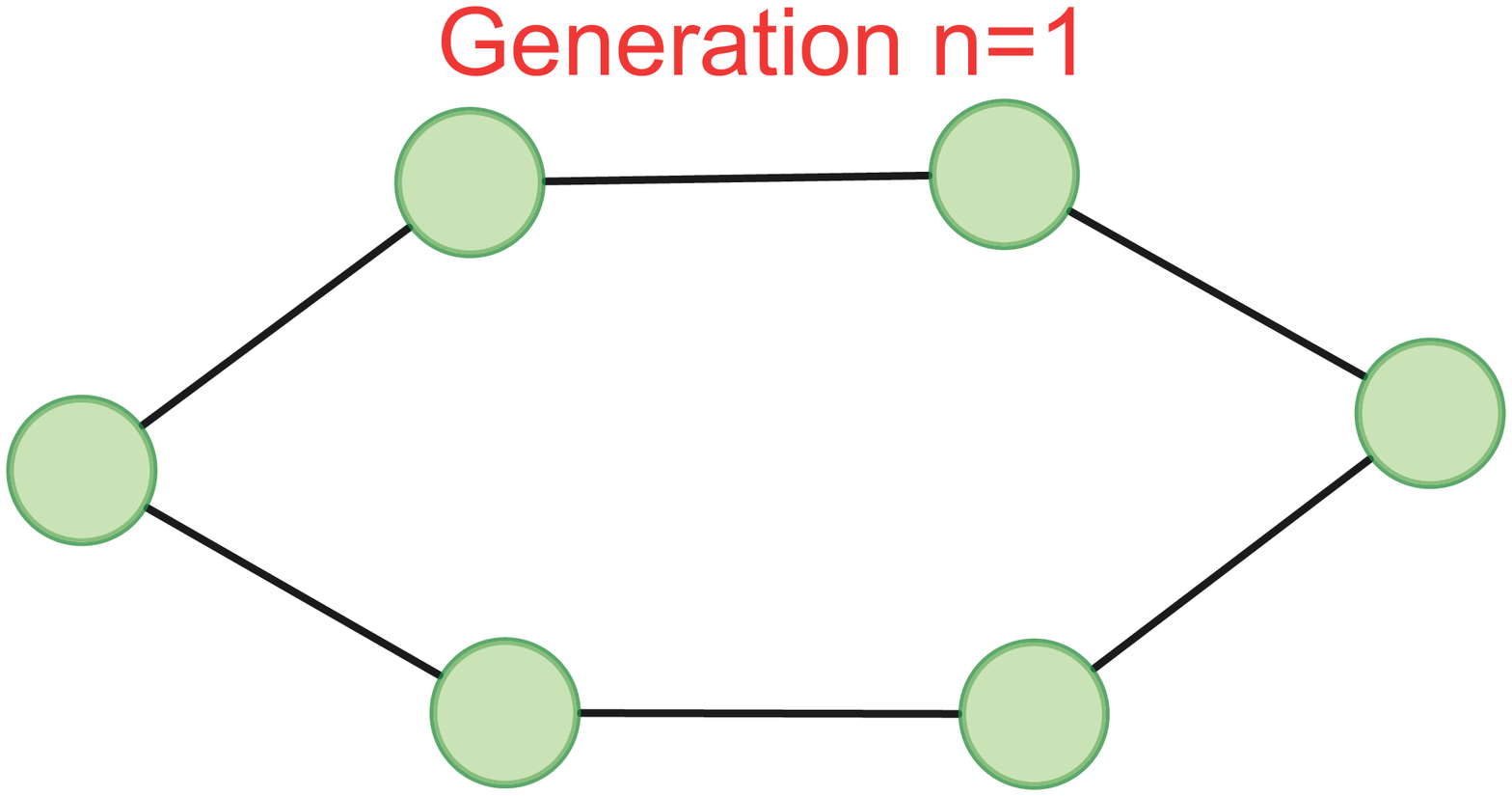} } \centerline{\epsfxsize=5cm
\epsfbox{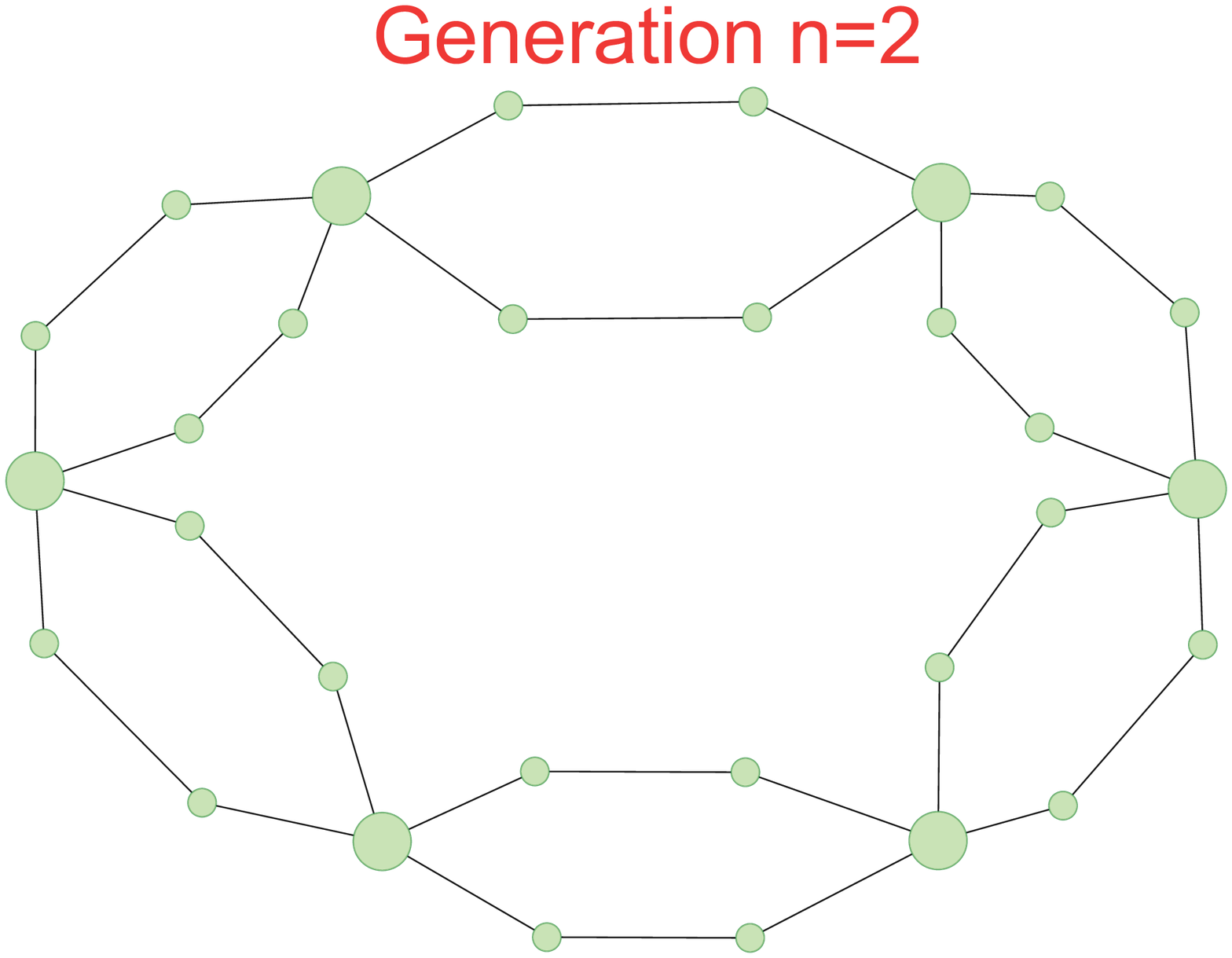}} \caption{Construction of a pure fractal
network. Example of the network modelof generations $n=0,1,2$ with
parameters $m=2,x=2,e=0$.}\label{2}
\end{figure}

According to above description, if we denote $M_n$ the number of
edges in the network of the generation $n$, we can have
$M_{n+1}=(2m+x)*M_n$ in the generation $n+1$. Hence ${{M_n}} =
{(2m + x)^n}$. Meanwhile in the growth of the network from the
generation $n-1$ to the generation $n$, each edge in the network
of the generation $n-1$ produces $2m$ new nodes. Hence we have
$N_n=2mM_{n-1}+N_{n-1}$, where $N_n$ is the number of nodes in the
network of the generation $n$. Therefore,
\begin{equation}
{N_n} = \frac{2m}{2m + x - 1}(2m + x)^n + 2 - \frac{2m}{2m + x -
1}
\end{equation}

It was proved in Song {\it et al.}~\cite{Song2006} that the degree
distribution $P(k)$ of the network model satisfies a power law
relationship $P(k) \approx {k^{ - \gamma }}$ with $\gamma=1+\ln
b/\ln s$, where $b$ is the scaling of the node number and $s$ is
the scaling of the node degrees between two adjacent generations
in the process of the growth of the network. From the algorithm
described above, we know that if the degree of a node in the
network of the generation $n$ is $K_n$, then it should be
$mK_n+K_n$ with probability $e$, or $mK_n$ with probability $1-e$
in generation $n+1$, therefore
$K_{n+1}=e(mK_n+K_n)+(1-e)mK_n=(m+e)K_n$. It is easy to know that
$b=2m+x$ from Eq. (1). So for the above model, we have
\begin{equation}
\gamma = 1 + \frac{{\ln (2m + x)}}{{\ln (m + e)}}
\end{equation}

In addition, a simple analysis shows that the clustering
coefficient of the network model is $0$ for any value of $e$.

\section{The fractal dimension}

\ \ \ \ We find that if the distance between two nodes of the
generation $n$ is $L_n$, then in the network of the generation
$n+1$, the distance $L_{n+1}$ would be $L_n$ with probability $e$,
or $3L_n$ with probability $1-e$. Hence
$L_{n+1}=eL_n+3(1-e)L_n=(3-2e)L_n$. From Ref. \cite{Song2006}, we
know that the theoretical fractal dimension of the model networks
is ${d_{f}^T}=\ln b/\ln a$, where $a=L_{n+1}/L_n$. Therefore,

\begin{equation}
{d_f^T} = \frac{{\ln (2m + x)}}{{\ln (3 - 2e)}}
\end{equation}
If $x=2,m=2$, we have $d_f^T=\ln (6) / \ln(3-2e)$.

We can also numerically calculate the fractal dimension of the
model networks using some algorithms (e.g. \cite{Song2007,
Kim2007}). Here we adopt the random sequential box-counting
algorithm proposed by Kim {\it et al.} \cite{Kim2007} to estimate
the fractal dimension of networks (two examples for estimating
fractal dimension are shown in Fig. 3). We denote $d_f^N$ the
fractal dimension of the network obtained numerically. First we
want to check whether values of $d_f^N$ coincide with the
theoretical values of fractal dimension $d_f^T$. If the numerical
and theoretical fractal dimensions coincide with each other, we
will have confidence on our process and program to estimate the
multifractal curves $\tau(q)$ and $D(q)$ of these networks. Due to
the limit of computational capacity of our computers, we only
generate the networks up to the generation $n=5$.

For each value of $e$, we generate 100 networks (100 realizations)
and calculate the average value of $d_f^N$ over the 100
realizations. The $e$ vs $<d_f^N>$ plot is presented in Fig. 4.
From Fig. 4, we can see that the numerical $<d_f^N>$ coincides
with the theoretical $<d_f^T>$ perfectly.

\begin{figure}[tbp]
\centerline{\epsfxsize=8cm \epsfbox{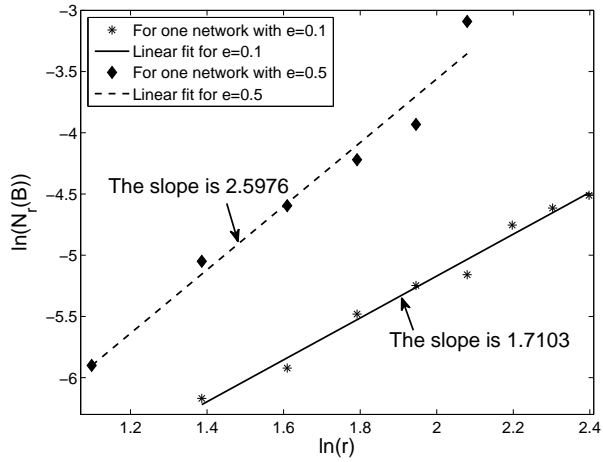}} \caption{
Two examples to estimate the fractal dimension of networks for
$e=0.1$ and $0.5$, here parameters $n=5,m=2,x=2$. We can see the
estimated fractal dimension is very close to the theoretical
result 2.5850 (for $e=0.5$) and 1.7402 (for $e=0.1$)
respectively.}
\end{figure}

\begin{figure}[tbp]
\centerline{\epsfxsize=8cm \epsfbox{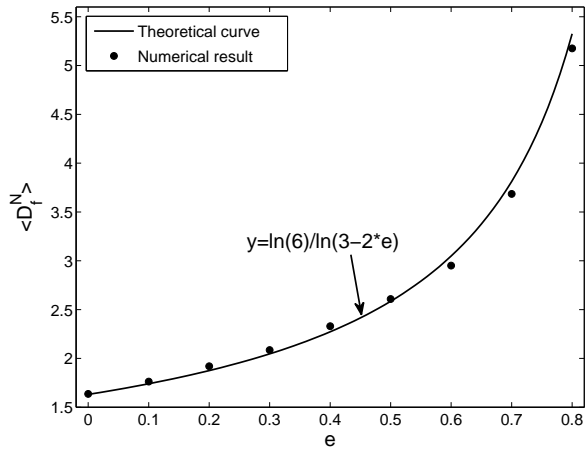}}
\caption{Numerical result of the relationship between the fractal
dimension $<d_f^N>$ of the networks and $e$ with parameters
$n=5,m=2,x=2$. Here $<\cdot>$ means the average over 100
realizations.}\label{3}
\end{figure}

\section{Multifractal analysis}
\ \ \ \ In this section, we first introduce a new algorithm for
MFA of networks modified from the one proposed in Ref.
\cite{Wang2012} and then apply it to the model networks presented
in Section 2.

Two networks which have the same fractal dimension may look
completely different. In addition, when the networks have rich
scale and self-similar structures, they exhibit different
dimensions in different scales. MFA is a powerful method to study
the networks with such characteristics.

At present, the fixed-size box-counting algorithm is the most
common algorithm for MFA \cite{Halsey1986}. For a given
probability measure $0 \le \mu  \le 1$ with support set $E$ in
metric space, we consider the partition function
\begin{equation}
{Z_r}(q) = \sum\limits_{\mu (B) \ne 0} {{{[\mu (B)]}^q}}
\end{equation}
where $q \in R$ , the result is the sum of all the different
non-overlapping boxes $B$ with a given size $r$ in the covering of
the support set $E$. It is easy to know that ${Z_r}(q) \ge 0$ and
${Z_r}(0) = 1$. We define the mass exponent function $\tau (q)$ of
the measure $\mu $ as
\begin{equation}
\tau (q) = \mathop {\lim }\limits_{r  \to 0} \frac{{\ln
{Z_r}(q)}}{{\ln r}}
\end{equation}

Then we get the generalized fractal dimensions of the measure
$\mu$ by
\begin{equation}
{D(q)} = \frac{{\tau (q)}}{{q - 1}},q \ne 1
\end{equation}
and
\begin{equation}
{D(1)} = \mathop {\lim }\limits_{r  \to 0}
\frac{{{Z_{(1,r)}}}}{{\ln r}},q = 1
\end{equation}
where${Z_{(1,r)}} = \sum\limits_{\mu (B) \ne 0} {\mu (B)\ln \mu (B)} $.\\

In practice, the generalized fractal dimensions are usually
obtained by linear regression. Specifically, $D(0)$ is the fractal
dimension of the support set of the measure $\mu$, ${D(1)}$ and
${D(2)}$ are called the information dimension and the correlation
dimension respectively.

For a network, the measure $\mu$ of each box can be defined as the
ratio of the number of nodes covered by the box to the total
number of nodes in the network~\cite{Wang2012, Kim2007}.  We need
to complete the following two steps before we proceed MFA.

i) Map a network to an adjacent matrix ${A_{N \times N}}$, where
$N$ is the total number of nodes in the network. It is easy to
know that ${A_{N \times N}}$ is a symmetric matrix where the
elements ${a_{ij}} = 1$ when there is an edge between the nodes
$i$ and $j$, otherwise ${a_{ij}} = 0$. Here, the edge from node
$i$ to node $i$ is not considered, so ${a_{ii}} = 0$.

ii) Use ${A_{N \times N}}$ to calculate the shortest distance
between any two nodes in the network and store them into another
matrix ${B_{N \times N}}$. Here, in our study, we use Dijkstra¡¯s
algorithm of MatLab toolbox to calculate the shortest distance
between two nodes in the network.

After finishing the two steps presented above, we can use matrix
${B_{N \times N}}$ as the input of MFA of the network model
described in Section 2 based on a new algorithm for MFA of
networks modified from the one proposed in Ref. \cite{Wang2012} as
follows.

\vspace{0.3cm} (I) Ensure that all nodes in the network are not
covered, and no node has been selected as the center of a box.

(II) According to the size of our networks $N=6222$ (with the
parameters $n=5;m=2,x=2$), we set $t = 1,2, \ldots ,T$. Here we
take $T=1000$, then we rearrange the nodes number into $T=1000$
different random orders. That is to ensure that the nodes of a
network are randomly chosen as center nodes.

(III) Set the radius $r$ of boxes which will be used to cover the
nodes in the range $[1,d]$, where $d$ is the diameter of the
network (i.e. the longest distance between nodes in the network).

(IV) Treat the nodes of the $t$th kind of random orders that we
have got in (II) as the center of a box successively, then search
all the other nodes. If a node has a distance to the center node
within $r$ and has not been covered yet, then cover it.

(V) If no more new nodes can be covered by this box, then we
abandon this box.

(VI) Repeat (IV) - (VI) until all the nodes are covered by the
corresponding boxes. We denote the number of boxes in this box
covering as $N(t,r)$.

(VII) Repeat steps (III) and (VI) for all the random orders to
find a box covering with minimal number of boxes $N(t,r)$.

(VIII) For each nonempty box $B$ in the first box covering with
minimal number of boxes, we define its measure as ${\mu (B) =
{N_B}/6222}$, where ${N_B}$ is the number of nodes covered by the
box $B$. For each $r$, we calculate the partition sum $Z_r(q) =
\sum\limits_{\mu (B) \ne 0} {{{[\mu (B)]}^q}}$.

(IX) For different $r$, we repeat (III)-(VIII). Then we use
$Z_r(q)$ for linear regression. \vspace{0.2cm}

{\bf Remark 1:} In the algorithm of MFA of networks proposed in
Ref. \cite{Wang2012}, we use $\overline{Z}_r(q)$ (the average of
$Z_r(q)$ for all $T=1000$ different random orders of the nodes)
for linear regression to get $\tau(q)$ (hence $D(q)$). But when
$q=0$, $D(0)$ got in this way is not the box-counting dimension of
the network because there requires minimum number of boxes which
cover the fractal set (network here) \cite{Falconer1997}. Here we
modify to use $Z_r(q)$ of a covering with minimum number of boxes
for linear regression to get $\tau(q)$ (hence $D(q)$). So when
$q=0$, $D(0)$ is exact the box-count dimension of the network. It
is a more reasonable extension from the traditional MFA.

 In order to get the range $r \in [{r_{\min
,}}{r_{\max }}]$ in which the networks obey the power law and then
to get the mass exponents $\tau(q)$ and the generalized fractal
dimensions $D_q$, linear regression is an important step. In our
calculation, we run the linear regression of $\ln {Z_r}(q)$
against $\ln r$ to get $\tau(q)$, and then get $D(q)$ through
formula $D(q)=\tau(q)/(q-1)$ for $q\ne 1$ and $D(1)$ through the
linear regression of $Z_{(1,r)} = \sum\limits_{\mu (B) \ne 0} {\mu
(B)\ln \mu (B)}$ against $\ln r$ for $q = 1$.

By applying the new fixed-size box-counting algorithm described
above on the model networks, we get the following results:

First, for each value of $e$ (here we take $e=0.1,0.2,...,0.8$),
we generate 100 networks (we take 100 realizations because the MFA
for networks is very time consuming when the network is large),
and calculate the $\tau(q)$ and $D(q)$ curves for each network
using the new fixed-size box-counting algorithm. Then we take
average for these $\tau(q)$ and $D(q)$ curves over the 100
realizations. The shape of the $<\tau(q)>$ curves shown in Fig. 5
and the $<D_q>$ curves shown in Fig. 6 are all nonlinear, which
indicate that all the networks we studied have multifractal
property. We also find that the value of $\Delta (<D(q)>)$ defined
by $\max (<D(q)>) -\min (<D(q)>)$ increases with the increase of
the parameter $e$, which indicates that the multifractal property
of the model networks becomes more obvious when the value of the
parameter $e$ becomes larger.

The multifractal property of the model networks revealed by our
work indicates that the model networks are very complicated which
cannot be characterized by a single fractal dimension. The MFA
algorithm proposed here can be used to provide a more accurate
characterization for the model networks, even for some other
complicated networks.

Second, we find that the average information dimension $<D(1)>$
has a linear relation with the parameter $e$, i.e.
$<D(1)>=1.5053*e+1.4735$ as shown in Fig. 7, which is different
from that of $D(0)$ shown in Eq. (3).

\begin{figure}[tbp]
\centerline{\epsfxsize=11cm \epsfbox{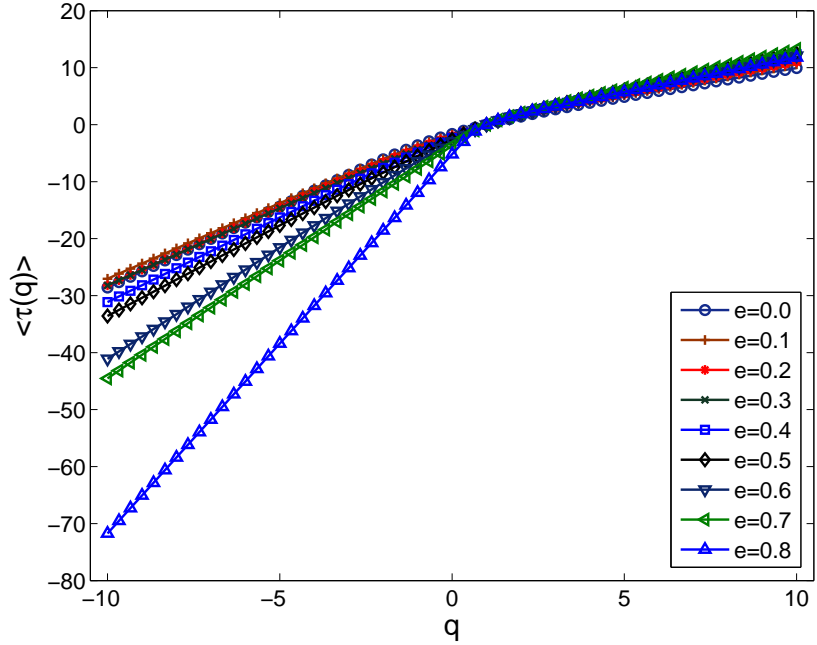}} \caption{The
$\tau(q)$ curves of the network model, here $<\cdot>$ means the
average over 100 realizations.}\label{5}
\end{figure}

\begin{figure}[tbp]
\centerline{\epsfxsize=11cm \epsfbox{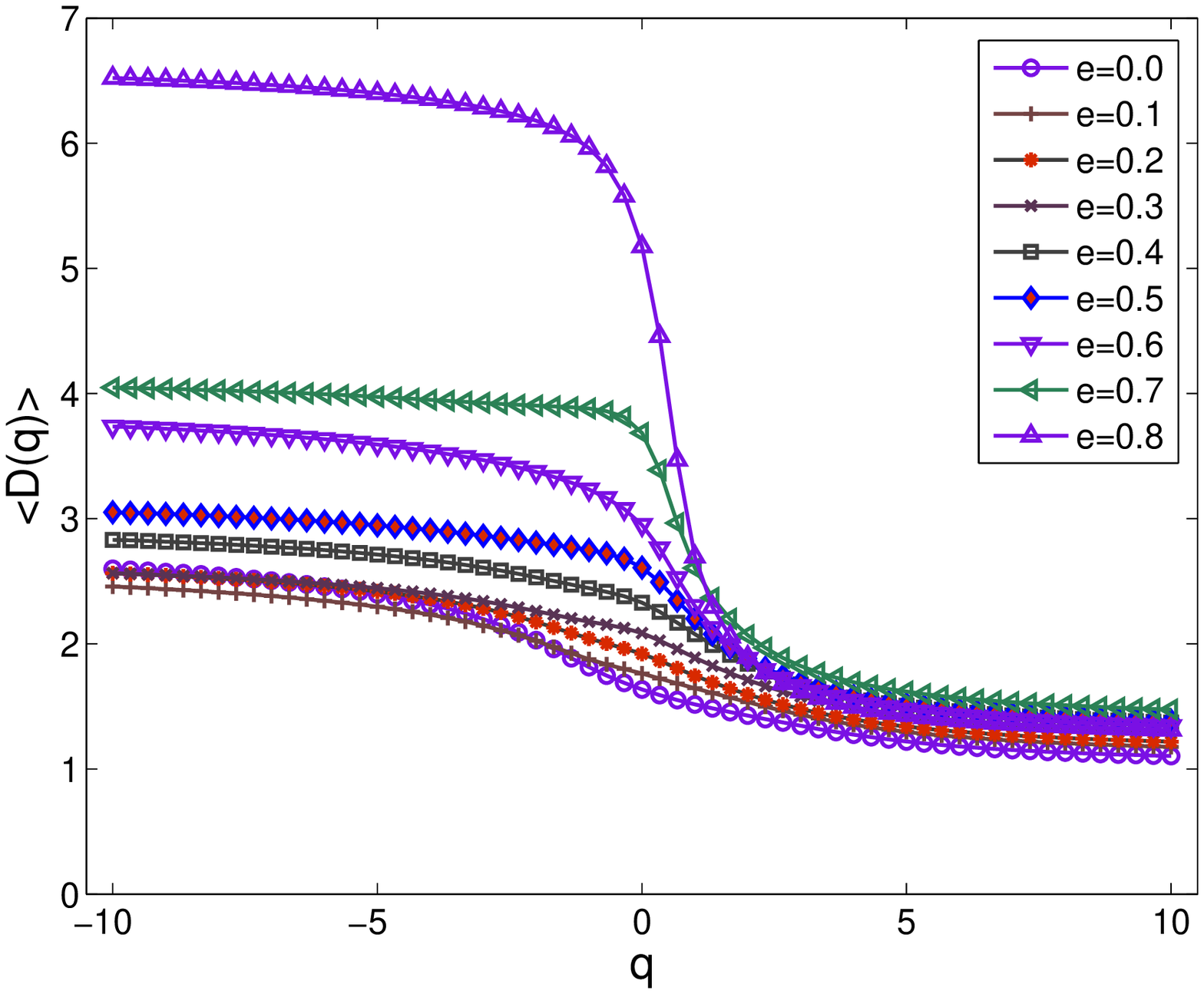}} \caption{The
$D(q)$ curves of the network model, here $<\cdot>$ means the
average over 100 realizations.}\label{5}
\end{figure}

\begin{figure}[tbp]
\centerline{\epsfxsize=11cm \epsfbox{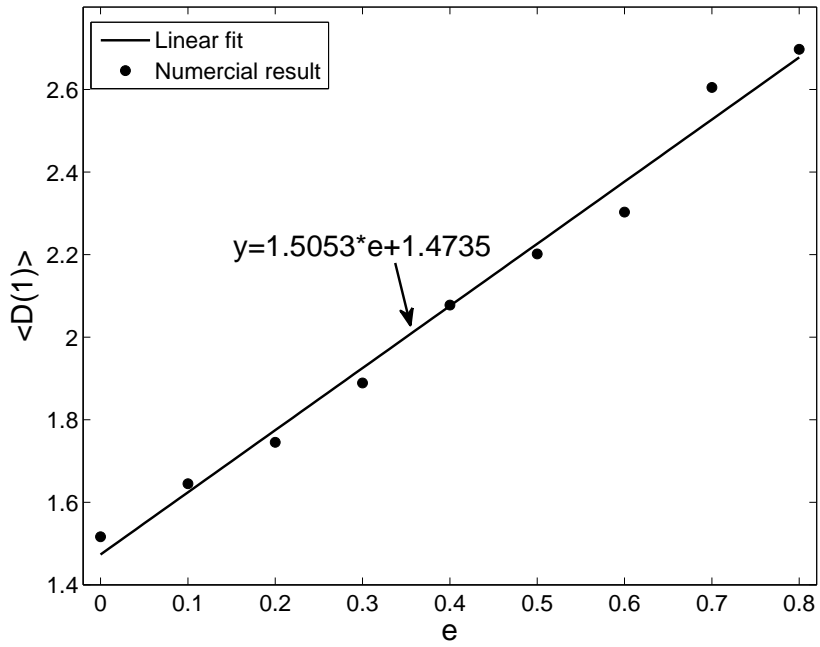}} \caption{The
relationship of $<D(1)>$ against parameter $e$ respectively, here
$<\cdot>$ means the average over 100 realizations.}
\end{figure}

{\bf Remark 2:} Furuya and Yakubo \cite{Furuya2011} also proposed
an algorithm for MFA of complex networks. The difference between
the algorithm in Ref. \cite{Furuya2011} and our algorithm is the
definition of the measure $\mu$. In the algorithm in Ref.
\cite{Furuya2011}, it allows that any two boxes in the box
covering have overlap and defines the measure $\mu_i$ by counting
the times of overlaps of each node, hence it is not a natural
extension of the traditional MFA (see Eq. (4)). In our algorithm,
overlap of any two boxes in the box covering is not allowed, so it
is a natural extension of the traditional MFA. Our network model
with $e>0$ is different from the ($u,v$)-flower network model.
Only the network model with $e=0$ corresponds to the deterministic
($u,v$)-flower network model with $u=v=3$. Furuya and Yakubo
\cite{Furuya2011} also gave a theoretical formula for the
$\tau(q)$ function of ($u,v$)-flower network model (Eq. (11) of
Ref. \cite{Furuya2011}). When $u=v=3$, $\tau(q)$ has the formula
\cite{Furuya2011}: $\tau(q)=q$ if $q\ge \ln (6) /\ln (2)=2.5850$,
and $\tau(q)=(q-1)\frac{\ln (6)}{\ln (3)}$ if $q<\ln (6) /\ln
(2)=2.5850$. We compared this formula with our numerical result
for $e=0$ in Fig.5, and found that we also have $\tau(q)\simeq q$
if $q\ge 2.5850$, but the $\tau(q)$ values are different from
$(q-1)\frac{\ln (6)}{\ln (3)}$ if $q<2.5850$.

\section{Conclusion}
\ \ \ \ We have studied the fractal and multifractal properties of
a family of model networks that were originally proposed to
explain the origin of fractality in complex networks. This model
introduces a parameter $e$, which can be used to tune the
fractality level of the network. One can get a pure fractal
network when $e=0$ and obtain a small-world network when $e=1$. We
investigated the fractal and multifractal properties through
numerical calculation. To make the calculation feasible and
accurate, we calculated the model with parameters ${n=5; m=2, x=2}
$; ${e=0}$, ${0.1, 0.2,\ldots,0.8}$; ${q=-10,\ldots, 10}$.  The
result of the $\tau(q),D(q)$ (including $D(0)$ and $D(1)$) are
averaged over $100$ realizations (networks). The shape of
$\tau(q)$ and $D(q)$ curves are all nonlinear, which indicates
that all the networks we studied have multifractal property. We
also found that the value of $\Delta (<D(q)>)=\max (<D(q)>) -\min
(<D(q)>)$ increases with the increase of the parameter $e$, which
indicates that the multifractal property of the model becomes more
obvious when the value of the parameter $e$ becomes larger.

We also found that the average information dimension $<D(1)>$ has
a linear relation with the parameter $e$, i.e.
$<D(1)>=1.5053*e+1.4735$.

The MFA algorithm proposed here can be used to provide a more
accurate characterization for the model networks, even for some
other complicated networks.

\section*{Acknowledgments}
\ \ \ \ This project was supported by the Natural Science
Foundation of China (Grant Nos. 11071282 and 11371016), the
Chinese Program for Changjiang Scholars and Innovative Research
Team in University (PCSIRT) (Grant No. IRT1179); the Research
Foundation of Education Commission of Hunan Province of China
(Grant No. 11A122); the Lotus Scholars Program of Hunan province
of China. The authors would like to thank the editor and the
reviewers for their insights, comments and suggestions to improve
this paper.

\end{document}